\documentclass[
fontsize=10pt,reprint,
superscriptaddress,
amsmath,amssymb,
aps,prd,floatfix
]{revtex4-2}

\usepackage{siunitx}
\newcommand{\e}{\epsilon}
\newcommand{\s}{\sigma}

\newcommand{\beq}{\begin{equation}}
\newcommand{\eeq}{\end{equation}}

\usepackage{graphicx}
\usepackage{dcolumn}
\usepackage{bm}

\usepackage[unicode=true,pdfusetitle,
   bookmarks=true, 
   bookmarksnumbered=false, 
   bookmarksopen=false,
   breaklinks=false, 
   pdfborder={0 0 1}, 
   backref=false, 
   colorlinks=true]{hyperref}

\hypersetup{pdfauthor={Name},colorlinks,
	linkcolor={blue},
	citecolor={blue},
	urlcolor={blue}}

\begin{document}

	\preprint{APS/123-QED}
	\sloppy
	\allowdisplaybreaks
	\title{Catalysis from the bottom-up}
	\author{Maitane Muñoz-Basagoiti}
	\affiliation{Gulliver UMR CNRS 7083, ESPCI Paris, Université PSL, 75005 Paris, France}
	\author{Olivier Rivoire}
 	\email{olivier.rivoire@college-de-france.fr}
	\affiliation{Center for Interdisciplinary Research in Biology (CIRB), Coll\`ege de France, CNRS, INSERM, Universit\'e PSL, 75005 Paris, France}
 	\author{Zorana Zeravcic}
    \email{zorana.zeravcic@espci.fr}
	\affiliation{Gulliver UMR CNRS 7083, ESPCI Paris, Université PSL, 75005 Paris, France}
\date{\today}

\begin{abstract}
Catalysis, the acceleration of chemical reactions by molecules that are not consumed in the process, is essential to living organisms but currently absent in physical systems that aspire to emulate biological functionalities with artificial components. Here we demonstrate how to design a catalyst using spherical building blocks interacting via programmable potentials, and show that a minimal catalyst design, a rigid dimer, can accelerate a ubiquitous elementary reaction, the cleaving of a bond. By combining coarse-grained molecular dynamics simulations and theory, and by comparing the mean reaction time in the presence and absence of the catalyst, we derive geometrical and physical constraints for its design and determine the reaction conditions under which catalysis emerges in the system. The framework and design rules that we introduce are general and can be applied to experimental systems on a wide range of scales, from micron size DNA-coated colloids to centimeter size magnetic handshake materials, opening the door to the realization of self-regulated artificial systems with bio-inspired functionalities.
\end{abstract}

\maketitle

\section{\label{sec:level1}Introduction}
Bio-inspired design combines principles from physics, chemistry, biology and engineering to create artificial materials with functionalities that rival those of biological systems, paving the way for the next generation of ``smart'' materials. One of the key ingredients for biological self-assembly and self-organization is the specificity of interactions between building blocks at the molecular scale. This has motivated an enormous experimental progress over the last decades in making artificial building blocks that differ in shape, size and types of interactions. For example, single stranded DNAs grafted on the surface of nano- and micron size particles lead to short-range binding specificity which controls what particle types can interact~\cite{Mirkin1996, Alivisatos1996, Rogers2011, Rogers2016,Elacqua2017, Cui2022}, patchy and asymmetric particles can be used for directional bonding with valence control~\cite{Feng2013Patchy, Hueckel2021}, and mobile DNA linkers on colloidal particles and emulsion droplets lead to valence control without predetermined particle geometry~\cite{Feng2013,Angioletti2014, McMullen2021, Chakraborty2022}. Following these advances, model systems based on such artificial building blocks have been used in experiments, theory and simulations to demonstrate desired properties like robust and reliable self-assembly into target structures~\cite{Rogers2016,Wang2017, He2020,Zhang2017, Zeravcic2014, Halverson2013,jacobsPNAS,Ong2017, Zion2017, Neophytou2021, McMullen2022}, structure reconfiguration~\cite{Zhang2019Dynamic,Rogers2015, Oh2020} and self-replication~\cite{Leunissen2009, Zeravcic2014a,Zhuo2019, Kim2015}. 
 
A major obstacle for efficiency and scalability in these artificial systems is the control over the formation and cleavage of targeted bonds. For instance, escaping kinetic traps, which are detrimental for the self-assembly yield of target structures~\cite{Grant2011,zeravcic2014size, Hedges2014, McMullen2022}, requires breaking bonds between particular building blocks. Likewise, in artificial self-replicating systems bonds between specific building blocks must successively form and break in a timely manner~\cite{Zeravcic2014a, Zhuo2019,Zhou2021}. The efficiency of these and many similar processes currently relies on external intervention, such as temperature and UV cycling protocols or mechanical forcing, which can non-specifically impact all the bonds in the system~\cite{Leunissen2009,Zhuo2019, Zhou2021, Schulman2012}. Nature, however, does it differently. In biological systems, reactions are facilitated by catalysts -- enzymes -- which, besides being extremely efficient and specific, are not energy consuming and are systematically recycled. Mimicking this level of control in artificial systems using designed building blocks will open the door to realizations of self-regulated systems with bio-inspired functionalities. 

One challenge is therefore to build tailor-made catalysts out of artificial building blocks. Over many decades, qualitative principles have been formulated, including notably Haldane’s principle of strain-based catalysis~\cite{haldane1930enzymes}, Pauling’s principle of complementarity to the transition state~\cite{pauling1946molecular} and Sabatier’s principle of ``just-right'' interaction strength~\cite{Sabatier1920}. These principles address different aspects of the design of a catalyst, but they do not constitute a general framework to build artificial catalysts from scratch for any desired reaction. As a result, successful designs of physical catalysts have so far relied on astute mechanisms that are not readily applicable beyond their original context~\cite{penrose1957self,miyashita2015mechanical, Zhang2007}, and only recently general considerations for catalysis have been investigated in an abstract model, shedding light on the constraints that apply to the design of a catalyst~\cite{rivoire2020geometry}. There is thus a need for empirical and theoretical insights with a bottom-up approach that integrates kinetic, geometrical and physical constraints to enable the design of catalysts that are experimentally implementable in physical systems. 

To make the first steps towards this goal, here we present the computational design of a minimal catalyst capable of accelerating the dissociation of a dimer into monomers, i.e., cleaving a bond. Starting from building blocks interacting via programmable potentials, we provide rules for making a catalyst and test our design in numerical simulations of a physical model system. To our knowledge, this is the first proposal of a physically realizable artificial catalyst that is designed from the bottom up. The framework and design rules we introduce are general and can be applied to experimental systems spanning different length scales, from micron size DNA-coated colloids~\cite{Rogers2016} to centimeter size magnetic handshake materials~\cite{niu2019magnetic}. 

\section{Model system}
\begin{figure}[t!]
\centering
\includegraphics[width=\columnwidth]{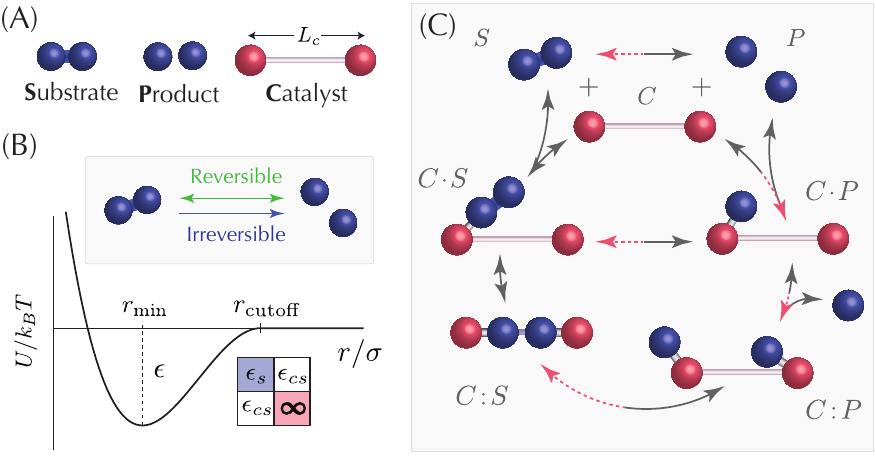}
\caption{\textbf{Model parameters.} {\bf(A)} Our system consists of a substrate dimer $S$ that can dissociate into two free product monomers $P$ (in blue), and a catalyst $C$ (in red). The particles in the catalyst are kept at a fixed distance $L_c$. {\bf (B)} We use a short-range pairwise interaction potential with depth $\epsilon$, interaction range $r_{\text{cutoff}}$ and equilibrium position $r_{\text{min}}$~\cite{Wang2020}. As indicated by the interaction matrix, blue particles interact with strength $\epsilon_{s}$, and blue and red particles interact with strength $\epsilon_{cs}$; the catalyst bond is rigid. The spontaneous reaction can be reversible or irreversible. {\bf (C)} Reaction scheme in the presence of the catalyst. Gray arrows correspond to the most favourable reaction conditions for catalysis: free monomers are removed from the system and the substrate bond is not allowed to reform once broken by the catalyst; dashed red arrows correspond to the worst-case conditions, where the above transitions are reversible and diffusion-limited.} 
\label{fig:model}
\end{figure}
\subsection{Spontaneous reaction}
The design of a catalyst depends on the reaction it accelerates. Here we focus on the dissociation of a dimer $S=M{\cdot}M$ into two free monomers $P=M\!+\!M$, which we refer to as the substrate $S$ and the product $P$ of the reaction (Fig.~\ref{fig:model}A). The substrate dimer is composed of two spherical particles of diameter $\s$ interacting via an isotropic pairwise potential of depth $\e_{s}$ and interaction range $r_{\text{cutoff}}$~\cite{Wang2020} (Methods), which, if exceeded, leads to the dissociation of the dimer into monomers (Fig.~\ref{fig:model}B). Inspired by the short-range interactions between colloids mediated by DNA (through direct hybridization or by using linkers)~\cite{Rogers2016, Rogers2020}, we set the interaction cutoff to be $r_{\text{cutoff}}=1.10~\sigma$, which results in $r_{\text{min}}=1.03~\sigma$ as the equilibrium position of the substrate bond (Methods). We consider two limiting cases for the reaction between the free monomers: irreversible, when the monomers cannot reform a bond and, reversible, when the bond formation is diffusion-limited. In what follows, we take $k_BT=1$ for the energy scale and $\s=1$ for the length scale in our system. The transition from the dimer to the two-monomer state occurs spontaneously through thermal activation. 
We perform Molecular Dynamics (MD) simulations of this system (Methods) and verify that the dissociation events, defined as the \textit{first time} the distance between two monomers exceeds $r>r_{\text{cutoff}}$, are exponentially distributed with mean reaction time $T_{S\to P}$ (Fig.~S1). This mean time is the key parameter that we will use for the assessment of catalysis.

\subsection{Catalyst design}
To build a minimal catalyst we use the same-sized spherical particles that comprise the substrate. As we verify in numerical simulations, a single particle cannot catalyze the dissociation reaction (Fig.~S2). The next simplest design is a dimer made of two particles whose centers are held at a distance $L_c$. Motivated by previous theoretical results~\cite{rivoire2020geometry}, we set the catalyst bond to be rigid. Each catalyst particle can interact with a substrate particle via the potential in Fig.~\ref{fig:model}B~\cite{Wang2020}, with a strength $\e_{cs}$ and an interaction range that we constrain to be the same $r_{\text{cutoff}}$. Bond formation between catalyst and substrate particles is diffusion-limited. For simplicity, we assume that particles exhibit valence: a catalyst particle can interact with only one substrate particle at a time, while a substrate particle can interact with only one substrate and only one catalyst particle at the same time. This type of restricted binding has been achieved in experiments with colloidal particles~\cite{Wang2012,Zhang2018,McMullen2021}, while the fixed distance $L_c$ between catalyst particles could be realized experimentally by placing them on a surface.

\subsection{System parameters}
Catalysis depends on three sets of parameters. First, it depends on the reaction to be accelerated, characterized by the overall shape of the interaction potential, and the entropic barriers for the forward and the reverse direction. In our simple model, these are controlled by the strength of the bond we want to cleave, $\e_s$, and by the (ir)reversibility of the spontaneous reaction respectively. Second, catalysis depends on the intrinsic properties of the catalyst, like its geometry and types of interactions. These correspond to the two particles rigidly bound, placed at a fixed distance $L_c$ and the interaction strength towards the substrate $\e_{cs}$ in our model. Finally, catalysis depends on extrinsic properties of the system such as the concentration of substrate and product and their diffusion constant, as well as the volume and temperature of the system, which we will discuss in more detail elsewhere~\cite{Yann2022th}.

Our goal here is to identify intrinsic parameters of the catalyst that lead to catalysis in the system. To assess if there is catalysis, we consider a box of volume $V$ with a single substrate, and compare the mean reaction time in the absence and presence of a single catalyst in the system, i.e., compare the mean first-passage times $T_{S\to P}$ to $T_{C+S\to C+P}$, where $C+S$ and $C+P$ account for the substrate and product particles in the presence of, but not interacting with the catalyst. A successful catalytic design must reduce the mean reaction time, and our criterion for catalysis is $T_{S\to P}/T_{C+S \to C+P}>1$. To proceed, we first explore the parameter space under two assumptions that are most favorable for catalysis, and then investigate what happens when we lift those constraints. The first assumption is that the spontaneous reaction is irreversible, preventing spontaneous formation of substrate bonds once broken. The second assumption is that the product monomers are removed from the system as soon as they are released in solution, preventing them from binding the catalyst. In the setup shown in Fig.~\ref{fig:model}C, red arrows correspond to the four backward processes that are excluded due to these assumptions.

\section{Results}
\subsection{Conditions for catalysis}
In order to identify necessary conditions for catalysis, we decompose the catalytic cycle into elementary processes, each of which corresponds to the formation or cleavage of a single bond. As illustrated in Fig.~\ref{fig:model}C, this defines six possible states of the system: $C+S$, where the substrate and catalyst are not interacting; $C{\cdot}S$, where the substrate forms one bond with the catalyst; $C{:}S$, where both substrate particles are bound to the catalyst; $C{:}P$, where the substrate bond has been cleaved and both products remain bound; $C{\cdot}P$, where one product particle has been released into the solution and the second one remains bound; and finally $C+P$, where the two product particles are released by the catalyst, which recovers its initial state. The reaction $C{\cdot}S\to C{\cdot}P$ corresponds to the spontaneous dissociation of the substrate bond while the substrate is partially bound to the catalyst. The series of reactions from state $C{:}S$ up until the recycling of the catalyst in $C+P$ comprise an alternate pathway to reach the final product state exclusively due to the presence of the catalyst.

A trivial necessary condition for catalysis is that the state $C{:}S$ is accessible from $C{\cdot}S$. Otherwise, the only way for product particles to appear in solution is through the transitions $S\to P$ and $C{\cdot}S\to C{\cdot}P$, i.e., through the spontaneous reaction. This condition constrains the distance between the catalyst particles $L_c$ to be small enough and imposes favorable binding between the catalyst and the substrate, i.e., $L_c<3r_{\text{cutoff}}$ and $\e_{cs}>0$, respectively.

Next, one expects the transition from $C{:}S$ to $C+P$ to take on average less time than the spontaneous reaction $S\to P$. This implies necessary conditions on the elementary steps (i)~$C{:}S\to C{:}P$, (ii)~$C{:}P\to C{\cdot}P$ and (iii)~$C{\cdot}P\to C+P$.
Process (i) leads to the necessary condition $T_{C{:}S\to C{:}P\backslash C{\cdot}S}<T_{S\to P}$, where $T_{C{:}S\to C{:}P\backslash C{\cdot}S}$ denotes the mean first-passage time from $C{:}S$ to $C{:}P$ excluding the possibility of the back transition to the state $C{\cdot}S$. In other words, cleaving the substrate bond in the presence of the catalyst should be faster than in its absence. To find which parameters this condition constrains, we perform MD simulations of one catalyst and one substrate in a box initiated in $C{:}S$ configuration, with fixed $\e_s$, and $\e_{cs}\gg 1$ to avoid substrate unbinding (Fig.~S2). By varying the geometry of the catalyst we find that the condition can be satisfied only if $L_c > 3r_{\text{min}}$. At $L_c= 3r_{\text{min}}$, which corresponds to the geometrical threshold above which the substrate can fit between the particles in the catalyst, the $C{:}S$ configuration is one-dimensional (see Fig.~\ref{fig:model}C) and all the bonds are in equilibrium. For larger $L_c$, the bonds are stressed, leading to the substrate bond being \textit{strained} by the catalyst.

Steps (ii) and (iii) lead to the additional necessary conditions $T_{C{:}P\to C{\cdot}P\backslash C{:}S}\!<\!T_{S\to P}$ and $T_{C{\cdot}P\to C+P\backslash C{:}P}\!<\!T_{S\to P}$. Given that these two steps correspond to the same process (breaking an $\epsilon_{cs}$ bond with the catalyst), and that the catalyst releases product monomers independently, both conditions are satisfied if $T_{C{\cdot}P\to C+P\backslash C{:}P}\!<\!T_{S\to P}$. Simply, releasing one product particle should be faster than the spontaneous reaction. This condition depends only on how strongly the product is bound to the catalyst, and is satisfied by imposing $\e_{cs}\!<\!\e_s$, i.e., the interaction between $S$ and $C$ should be weaker than the scissile bond in the substrate (Methods).

No similar conditions apply to the remaining two elementary steps towards $C+P$ in Fig.~\ref{fig:model}C, namely, transitions $C+S\to C{\cdot}S$ and $C{\cdot}S\to C{:}S$. First, the emergence of catalysis is independent of the time it takes the substrate to find the catalyst $T_{C+S\to C{\cdot}S}$, because the spontaneous dissociation of the substrate into two monomers can always occur along the way. Second, the necessary condition on $T_{C{\cdot}S\to C{:}S}$ non-trivially involves mean reaction times of other elementary steps in the catalytic pathway because of the $C{\cdot}S \to C{\cdot}P$ transition. As a result, no simple additional constraint on the catalyst design can be derived. We discuss these transitions in more details in the SI and more formally in~\cite{Yann2022th}.

\subsection{Phase diagram for catalysis}
\begin{figure}[t!]
\includegraphics[width=\linewidth]{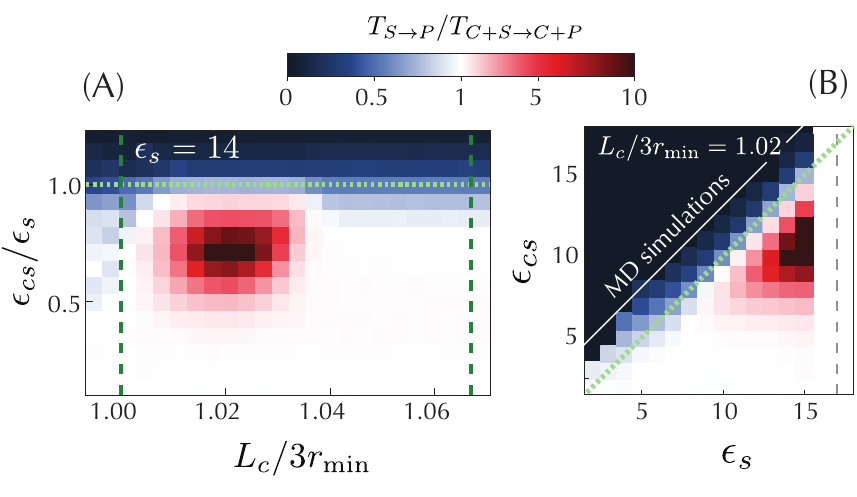}
\caption{\textbf{Phase diagram of catalysis.} Simulation results for a 2D system under conditions that are most favorable for catalysis, i.e., systematic product removal and irreversible spontaneous reaction. The simulation box length is $L/\sigma\!=\!7.5$. Event statistics over which the average reaction times are computed are shown in Fig.~S3. {\bf (A)} For a fixed spontaneous reaction ($\epsilon_s= 14$) in the irreversible limit, catalysis, i.e., $T_{S\to P}/T_{C+S\to C+P}\!>\!1$, requires $3 r_{\text{min}}\!<\!L_c\!<\!3 r_{\text{cutoff}}$ (green dashed lines) and $\epsilon_{cs}/\epsilon_s<1$ (green dotted line). Catalysis (in red) is indeed observed only within these bounds. {\bf (B)}\! Fixing the catalyst geometry ($L_c/3r_{\text{min}}\!=\!1.02$) shows that there is a minimal $\epsilon_s$ required for catalysis. The diagonal green dotted line represents the $\epsilon_{cs}/\epsilon_s\!<\!1$ constraint. The white solid line separates simulated data (below) from extrapolated data (above). The grey dashed line gives an indication of experimental time of $1$ \si{\second}  for a model system of colloids with $\sigma\!=\!1~\mu$m at room temperature.}
\label{fig:cond}
\end{figure}

For a given spontaneous reaction, i.e., for a given $\epsilon_s$ in the irreversible limit, and based on the above analysis, the necessary conditions for catalysis in our model constrain the design parameters for the dimer catalyst to $3r_{\text{min}}<L_c<3r_{\text{cutoff}}$ and $0<\epsilon_{cs}\!<\!\epsilon_s$. We perform MD simulations of the system with and without the catalyst present within this parameter range and compare the mean reaction times to produce two free monomers in both cases. Our results for a 2-dimensional (2D) system are shown in the phase diagrams in Fig.~\ref{fig:cond}. 

In Fig.~\ref{fig:cond}A we show catalytic efficiency, i.e., $T_{S\to P}/T_{C+S \to C+P}$, for different values of $L_c$ and $\epsilon_{cs}$, with $\epsilon_s$ fixed. Under these conditions, we demonstrate that catalysis can occur within the range of identified necessary conditions (dashed and dotted lines). The maximum in the heat map reveals the best design for the catalyst. In Fig.~\ref{fig:cond}B we now fix the catalyst geometry $L_c$ and vary $\epsilon_s$ and $\epsilon_{cs}$, showing again that catalysis occurs within the range of necessary conditions.

Our detailed simulations in the prescribed $(\e_{cs},L_c,\e_s)$ parameter space reveal that catalysis requires a minimal value of $\e_s$ (Fig.~\ref{fig:cond}B). This is another condition on catalysis that applies to the spontaneous reaction rather than to the catalyst design itself. We interpret this result through two arguments. Firstly, if the spontaneous reaction occurs too fast, the substrate will dissociate before it can bind the catalyst in the right configuration. Secondly, the catalyst inhibits the reaction for too small $\e_s$ because the substrate is able to diffuse to the catalyst, reaching the $C{\cdot}S$ state, but the bond breaks at its spontaneous rate along the $C{\cdot}S\to C{\cdot}P$ pathway instead of using the catalytic mechanism. One of the monomers is released in solution, while the second one remains attached on the surface of the catalyst. Since the reaction ends only once there are two free monomers in solution, and the catalytic pathway contains the extra step of release which does not exist for the spontaneous reaction, the design effectively inhibits the reaction by delaying the production of free monomers. As energies are here given in units of $k_BT$, this last condition indicates a threshold temperature above which catalysis does not occur anymore. 

\subsection{Trade-offs}
Our phase diagram in Fig.~\ref{fig:cond} reveals trade-offs applying to the geometry $L_c$ of the catalyst and how strongly it binds the substrate $\e_{cs}$. These trade-offs emerge from considering \textit{all elementary steps} in the catalytic cycle simultaneously.

In the case of the catalyst geometry, within the bounds $3r_{\text{min}}<L_c<3r_{\text{cutoff}}$, there is an optimum $L_c$ that maximizes the strain on the substrate bond while minimizing the time to fully bind the substrate. The strain is greater the more complementary the catalyst is to the transition state (Fig.~S2), in agreement with Pauling's principle~\cite{Pauling1946}, where exact complementarity in our model implies $L_c/3r_{\min}=2/3+r_{\rm cutoff}/3r_{\min}= 1.023$ (Methods). However, adopting this configuration takes long time (Fig.~S4), giving rise to the trade-off~\cite{Wolfenden1974}. In the case of the catalyst-substrate bond, the optimal strength $\e_{cs}$ maximizes the strain on the substrate while minimizing the time to release the product particles into solution, which embodies Sabatier's principle~\cite{Sabatier1920} of optimal intermediate binding strength. 

\begin{figure*}[!ht]
\centering
\includegraphics[width=\linewidth]{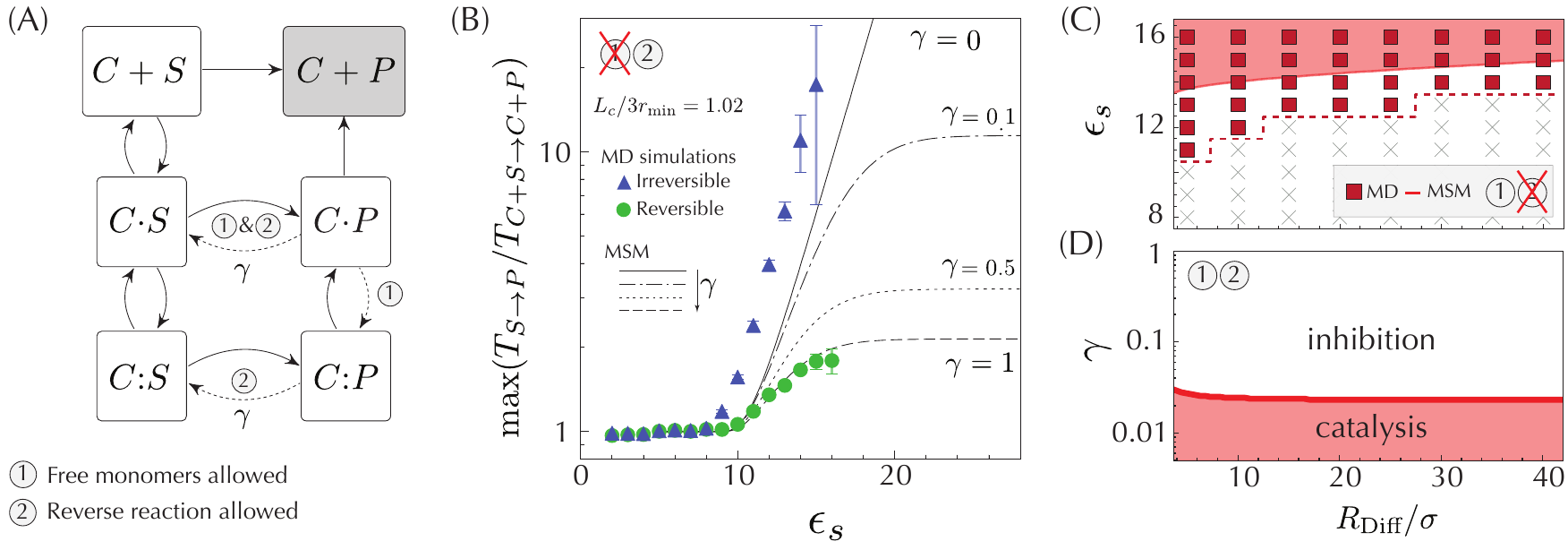}
\caption{\textbf{Lifting constraints in the model.} {\bf (A)} Structure of the Markov State Model (MSM) inferred from MD simulations showing the states of the system in the presence of the catalyst. The reaction is over when state $C+P$, colored in gray, is reached. We consider two different constraints on the model: (1) removing monomers as soon as they are released into solution; and (2) limiting the reformation of the substrate bond through the parameter $\gamma\in [0, 1]$, with $\gamma = 1$ corresponding to reversible reactions and $\gamma = 0$ to irreversible ones. Dashed arrows indicate the transitions affected by these constraints. {\bf (B)} Maximal catalyst efficiency $\max(T_{S\to P}/T_{C+S\to C+P})$ when free monomers are removed from the system as a function of the substrate bond strength $\epsilon_s$ for a fixed geometry of the catalyst $L_c/3r_{\min} = 1.02$. Data points correspond to 2D MD simulations and black lines to the MSM depicted in panel A. The maximal efficiency in the model, which is obtained when $\gamma = 0$, scales exponentially as $\alpha \epsilon_s$, with a factor $\alpha_{\text{MD}} = 0.50\pm 0.04$ (the fit is conducted for points where the efficiency is larger than one, i.e., $\epsilon_s \geq 8$) and $\alpha_{\text{MSM}} = 0.44$. The maximal efficiency saturates for large $\epsilon_s$ when $\gamma \neq 0$ (SI). {\bf (C)} Substrate bonds $\epsilon_s$ for which catalysis is observed (red) in a 2D system with free monomers removed only after they have diffused a distance $R_{\text{Diff}}/\sigma$ from the catalyst. We keep the spontaneous reaction irreversible, i.e., $\gamma = 0$. Red region represents results from the Markov model and red squares are results from our MD simulations. White region and gray crosses mark the regions where catalysis is not possible in the model and simulations respectively. {\bf (D)} MSM results showing the $\gamma$ values for which catalysis can be observed in 2D when monomers are removed from the system if they have diffused a distance $R_{\text{Diff}}/\sigma$ from the catalyst.}
\label{fig:efficiency}
\end{figure*}

\subsection{Coarse-graining and relaxing constraints}

To examine the conditions for catalysis beyond the restrictive assumptions that we made so far -- irreversible reaction and systematic removal of products -- we follow the approach usually taken in chemistry to coarse-grain chemical reactions as Markov processes~\cite{Husic2018}. Under this approximation, a larger range of parameters can be more efficiently explored. In general, one cannot assume that all the steps in a cycle (as in Fig.~\ref{fig:model}C) can be represented as Markov transitions. We verify, however, that this is the case in our model system within the range of parameters necessary for catalysis, which means that the rates for the transition between states can be inferred from our simulations, an approach previously applied in other MD studies~\cite{Chodera2014, Sarich2013, Malmstrom2014} (Fig.~S5). We use rates inferred from MD simulations to verify the coarse-graining (Methods), and the resulting Markov State Model (MSM) is shown in Fig.~\ref{fig:efficiency}A. To extend the exploration of the parameter space, we develop an analytical model for the dependency of the rates on the parameters (Methods) that we use in all MSM calculations that follow.

We first consider relaxing only the condition that the reaction is irreversible. In the MSM we therefore introduce a parameter $\gamma \in [0, 1]$ as a prefactor to the rates corresponding to diffusion-limited transitions $C{\cdot}P\to C{\cdot}S$ and $C{:}P\to C{:}S$. This allows us to interpolate between the reversible ($\gamma = 1$) and irreversible ($\gamma = 0$) limits of substrate bond formation (Fig~\ref{fig:efficiency}A). In Fig.~\ref{fig:efficiency}B we show maximal catalytic efficiency for increasing substrate bond strength $\epsilon_s$ at different values of the parameter $\gamma$ (black lines). For $\gamma = 0$, the maximal efficiency of our catalyst scales exponentially with $\e_s$ (SI). When $\gamma > 0$, the transition $C{:}P \to C{:}S$ is possible. As a result, an additional constraint on catalysis arises, which couples the cleavage of the substrate bond and release of the first product monomer to the reformation of the substrate bond. This condition is responsible for the saturation of the catalytic efficiency at high $\epsilon_s$ seen in Fig.~\ref{fig:efficiency}B (SI). These Markov state model results agree with our MD simulations in the irreversible and reversible limit, shown as blue triangles and green circles in the figure, validating the coarse-graining of the catalytic pathway into states (see SI for additional validations).

Next we explore whether we can measure catalytic activity of our dimer catalyst if free monomers are not taken out of the system as soon as they are released into the solution. To do this, we introduce a disk of radius $R_{\text{Diff}}$ centered around the catalyst and consider that we have reached the $C+P$ state only once both free monomers diffuse out of this volume (Fig.~S7). Results we obtain from  the MSM when $\gamma = 0$ are shown in Fig.~\ref{fig:efficiency}C (red shaded region). As can be seen, under these conditions, the onset of catalysis depends on the volume of the disk around the catalyst: the larger the volume, the longer it takes free monomers to diffuse out, which implies that the spontaneous reaction must be slower, i.e., the minimal $\epsilon_s$ should be larger to observe catalysis in our model. Our 2D MD simulations, reporting the minimal substrate bond strength at which we observe catalysis in the irreversible limit (red squares), agree with the Markov model predictions. 

Finally, when $\gamma > 0$, our MSM predicts catalysis in 2D to be possible only for a range of $\gamma<\gamma_{\max}$ values, as shown in Fig.~\ref{fig:efficiency}D. In other words, our model predicts that beyond the most favorable conditions, the reverse reaction cannot be diffusion-limited, and that some finite barrier is necessary in order to observe catalysis. To verify this prediction in MD simulations requires first changing how particles interact, i.e., introducing a finite backward reaction barrier, and second expanding the range of substrate bond strengths explored, which becomes computationally more challenging as $\epsilon_s$ grows. We leave this verification for future work. Results for a 3D system are shown in the SI (Fig.~S9).
 
\subsection{Efficiency and optimality of the catalyst design}
\begin{figure}[t!]
    \centering
    \includegraphics[width=\linewidth]{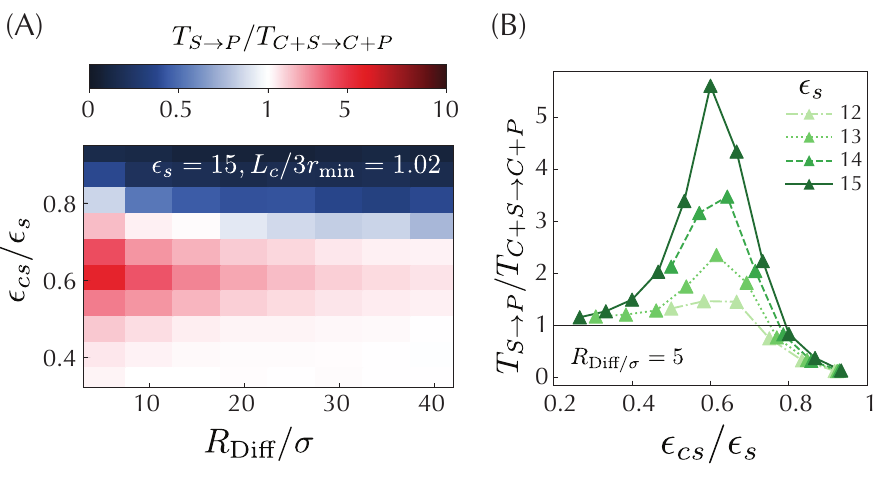}
    \caption{\textbf{Efficiency and catalyst design beyond most favourable reaction conditions.} MD simulation results in 2D when free monomers are removed if they diffuse beyond $r>R_{\text{Diff}}$, where $R_{\text{Diff}}/\sigma$ is the radius of a disk centered around the catalyst (Fig.~S7). The spontaneous reaction is assumed to be irreversible. {\bf (A)} Catalytic efficiency, $T_{S\to P}/T_{C+S \to C+P}$ for a fixed substrate bond with $\epsilon_s = 15$ and catalyst geometry $L_c/3r_{\min} = 1.02$, for different simulation box sizes and varying $\epsilon_{cs}$. {\bf (B)} Volcano plots showing the optimal catalyst binding strength $\epsilon_{cs}$ for a fixed volume $R_{\text{Diff}}/\sigma =5$ and varying substrate bonds. The optimal $\epsilon_{cs}/\epsilon_s$ remains unchanged as $\epsilon_s$ varies. The average for each data point is taken over at least 200 simulations. 
    }
    \label{fig:simulations}
\end{figure}

Motivated by possible experiments to be discussed below, we quantify the efficiency of our catalyst as function of bond strengths in the conditions where the product is not immediately removed, while the catalyst geometry ($L_c$) is kept fixed and the spontaneous reaction is irreversible. We first focus on the effect $R_{\text{Diff}}$ (volume) has on the maximal efficiency, which is represented by blue triangles in Fig.~\ref{fig:efficiency}B. While our simulation results in Fig.~\ref{fig:simulations}A show (as expected) that the maximal efficiency decreases with increasing volume (increasing $R_{\text{Diff}}/\sigma$), our minimal catalyst for a given spontaneous reaction (fixed $\epsilon_s$) still succeeds in accelerating the spontaneous reaction several fold and the maximal efficiency still scales exponentially in the limit $\epsilon_s \gg 1$. Moreover, our simulations demonstrate the robustness of our catalyst design. First, as seen in Fig.~\ref{fig:simulations}A, the optimum $\epsilon_{cs}/\epsilon_s \approx 0.6$ remains unchanged as the volume varies. Second, as seen in Fig.~\ref{fig:simulations}B, the optimum $\epsilon_{cs}/\epsilon_s$ remains unchanged also when $\epsilon_{s}$ varies for fixed $R_{\text{Diff}}/\sigma$. In other words, the curves of catalytic efficiency, known as Volcano plots in the catalysis literature~\cite{Balandin1969, Wodrich2021}, share approximately the same optimum $\epsilon_{cs}/\epsilon_s$.

\section{Discussion}
We have taken a first step toward introducing catalysis in artificial physical systems by presenting design rules for constructing the simplest possible catalyst $-$ a rigid dimer $-$ that can accelerate bond dissociation. Our design is directly implementable in a physically-realizable system of building blocks interacting via programmable potentials. Along the way, we have outlined a general approach to design catalysis from the bottom-up. Breaking up the catalytic cycle into steps allows us to derive necessary conditions that limit the region in the parameter space where catalysis can emerge. These conditions lead to trade-offs when simultaneously considered, as we verify in MD simulations. 

Our design rules are general and can be applied to a range of experimental systems with programmable interactions. Spherical particles with short-range specific interactions we focused on here can be easily realized with DNA-coated colloids~\cite{Rogers2016}. Particle valence, which we assumed in our model, is readily available in these systems. For example, emulsion droplets functionalized with mobile DNA strands can exhibit valence by controlling the strand density~\cite{McMullen2021}. Likewise, droplets can be functionalized with mobile DNA-origami constructs, where their number sets the valence~\cite{Zhang2018}. Controlling the geometry of the catalyst can simply be achieved by patterning a surface with precisely spaced DNA patches on which catalyst particles can be deposited. Alternatively, DNA-origami constructs could also be used to fix the distance between the particles in the catalyst~\cite{Hayakawa2022}, or the desired catalyst geometry could be 3D printed~\cite{Doherty2020} and functionalized with DNA afterwards. Finally, the most favourable reaction conditions in our model require controlling the reaction between two product monomers as well as the re-binding of products to the catalyst (product inhibition). Linker-mediated interactions~\cite{Rogers2020} can introduce an entropic barrier for the reformation of the substrate bond once broken, controlled by the concentration of the free floating linkers in solution. Similarly, self-protected attractions in DNA-functionalized particles could serve to minimize product inhibition~\cite{Leunissen2009}. Note that because our catalyst is robust, i.e., the optimum $\epsilon_{cs}/\epsilon_s$ remains unchanged when $\epsilon_{s}$ varies, the same substrate dimer and catalyst dimer could be used in experiments at a range of temperatures. 

Our catalyst operates through a strain mechanism first proposed by Haldane for enzymes~\cite{haldane1930enzymes}. This is however not the only possible mechanism of catalysis. In particular, other spontaneous reactions, such as bond formation, where the barrier to overcome is entropic rather than energetic, require different mechanisms which will be interesting to investigate in future work.

Bond cleavage plays a role in essentially all reactions. Our catalyst may therefore find applications in problems of self-assembly and self-replication where bond cleavage is currently non-specific and externally driven. A catalyst provides several advantages over such protocols: it can be made specific to a particular bond, it does not require intervention and energy input, and, as we have here demonstrated, it can be implemented using the same building blocks as the rest of the system. Although minimal, our catalyst provides insights into the design principles underlying catalysis, opening the door to a control over the reactions in bio-inspired artificial systems.

\vspace{1cm}
\noindent We would like to thank Jasna Brujic, Ludwik Leibler, Angus McMullen, Cl\'ement Nizak, Ben Rogers, and Yann Sakref for insightful discussions. This work has received funding from the European Union’s Horizon 2020 research and innovation program under the Marie Sklodowska-Curie grant agreement No.~754387. ZZ acknowledges funding from the city of Paris EMERGENCE(S) grant.

\section*{Methods}
\noindent \textbf{Model and numerical simulations.} We model the interactions between particles with an isotropic interaction potential introduced by Wang~\text{et al.} in~\cite{Wang2020}, where
\beq
U(r, \epsilon) = \e \alpha(r_{\text{cutoff}}, \sigma) \left[ \left( \frac{\sigma}{r} \right)^2-1 \right]\left[ \left( \frac{r_{\text{cutoff}}}{r} \right)^2 -1 \right]^ 2 
\eeq
if $r\!<\!r_{\text{cutoff}}$ and zero elsewhere, and where $\alpha(r_{\text{cutoff}}, \sigma)$ fixes $U(r_{\min}, \epsilon)=\e$. The equilibrium position $r_{\min}$ is given by the minimum of $U(r, \epsilon)$, which for $r_{\text{cutoff}}/\sigma = 1.1$ results in $r_{\min}/\sigma \approx 1.03$. When the distance between two particles that cannot interact due to valence restrictions is $r<r_{\min}$, they repel via a soft-harmonic potential,
\begin{equation}
    U_{H}(r) = \frac{1}{2} k (r-r_{\min})^2
\end{equation}
where we take $k = 1000$. We perform Langevin Molecular Dynamics (MD) simulations using an in-house code that implements a modified velocity-Verlet algorithm to integrate the equations of motion~\cite{Groot1999}. All simulations are performed with temperature $T = 1.0$, friction coefficient $\gamma = 12.5$ and time-step $\Delta t = 10^{-4}$ time units, and we used periodic boundary conditions. To investigate the effect of product inhibition (Fig.~\ref{fig:efficiency}C and Fig.~\ref{fig:simulations}), the side of the simulation box is chosen as $L = 2R_{\text{Diff}} + \sigma$, where $R_{\text{Diff}}$ is the radius of a disk centered around the catalyst  (Fig.~S7). The latter is placed in the center of the simulation box.

\noindent \textbf{Physical and geometrical constraints for catalysis.} 

\textit{The $\epsilon_{cs}<\epsilon_s$ condition}: MD results show that the mean first passage times for the $S\to P$ and $C{\cdot} P \to C+P $ transitions scale exponentially as $T_{S\to P} \propto e^{-A\epsilon_s}$ and $T_{C{\cdot}P\to C+P\backslash C{:}P} \propto e^{-A\epsilon_{cs}}$, with $A = 0.91$ (SI). Thus, $T_{C{\cdot}P\to C+P\backslash C{:}P}<T_{S\to P}$ leads to $\epsilon_{cs}<\epsilon_s$ for sufficiently large $\epsilon_s$.

\textit{Complementarity to the transition state}: When $L_c/3r_{\min}>1$, the $C{:}S$ configuration is linear, and it is not possible for all three bonds to adopt their equilibrium value (Fig.~S2B). As a consequence, the substrate bond is strained and the effective barrier for $C{:}S \to C{:}P$ is given by
\begin{multline}\label{equation}
    \Delta U_{C{:}S \to C{:}P} =   2U\left(\frac{L_c-r_{\text{cutoff}}}{2}, \epsilon_{cs}\right) - \\ \min_{r_s^*, r_{cs}^*} \Big [U(r_s,\epsilon_s) + U(r_{cs}, \epsilon_{cs}) \\ + U(L_c - r_s - r_{cs}, \epsilon_{cs}) \Big],
\end{multline} 
where $r_s$ represents the substrate bond length, $r_{cs}$ the substrate-catalyst bond length, the first term in the equation is the potential energy for $r_s = r_{\text{cutoff}}$, i.e., when the substrate is at the transition state, and the second term is the minimum of the potential energy when the substrate is completely attached to the catalyst. The same approach for the effective barrier for a one-dimensional catalyst has been reported in~\cite{rivoire2020geometry}. The smaller $\Delta U_{C{:}S \to C{:}P}$, the faster the substrate bond will break (Fig.~S2). The first term in~\eqref{equation} reaches its minimum when $L_c = 2r_{\min} + r_{\text{cutoff}}$. This threshold marks when the catalyst is geometrically completementary to the transition state.

\noindent \textbf{Construction of the Markov State Model.} We construct a MSM by coarse-graining the MD trajectories into discrete states (configurations in Fig.~\ref{fig:model}C). We first infer the transition rates from the discretized trajectories using $k_{ij} = p_{ij}/\tau_i$, where $k_{ij}$ is the rate from state $i$ to state $j$, $\tau_i$ is the average time the system stays at state $i$, and $p_{ij}$ is the jump probability from $i$ to $j$ \cite{Schutte2011}. To extract $\tau_i$ and $p_{ij}$ from simulations, we initiate the system in state $C+S$ and sample the system every $\tau_{\text{Lag}} = 50 $ time units recording the formation and breaking of bonds, where $\tau_{\text{Lag}}$ is chosen sufficiently large to ignore barrier recrossings (e.g. immediate reformation of a bond after breaking) while still allowing us to resolve the states along the catalytic pathway. This procedure leads to the merging of states $C{:}S$ and $C{:}P$ (Fig.~S5). Transition probabilities are then computed by measuring the average transition frequency between states \cite{Bhat1961}. 

To explore the parameter space beyond simulations, we next construct an analytical model for the rates by classifying the transitions in the MSM into escape and diffusive processes.

\textit{Escape processes}. We describe bond breaking events as barrier escape problems with Arrhenius-like expression for the rate of the transition, $k(\epsilon) = e^{-A \epsilon + B}$ with $A = 0.91$ and $B = 2.20$ (Fig~S1B). The broken bond for transitions $C{\cdot}S \to C+S $ and $C{\cdot}P \to C+P$ corresponds to $\epsilon_{cs}$, and hence, the rate is $k(\epsilon_{cs})$. The same bond is broken during the $C{:}P \to C{\cdot} P$ transition, but $k_{C{:P}\to C{\cdot}P} \approx 2 k_{C{\cdot}P \to C+P}$ as any of the two monomers attached to the catalyst can be released independently. The bond broken during the $C{\cdot}S \to C{\cdot}P$ transition is the scissile bond in the substrate, and therefore, the rate is $k(\epsilon_s)$. The barrier for $C{:}S \to C{:}P$ is described by~\eqref{equation}. The barrier for the $C{:}S\to C{\cdot}S$ transition is approximated by
\newpage
\begin{multline}
    \Delta U_{C{:}S \to C{\cdot} S}= U(r_{\min}, \epsilon_s) + U(r_{\min}, \epsilon_{cs}) \\
    - \min_{r_s^*, r_{cs}^*} \Big [U(r_s,\epsilon_s) + U(r_{cs}, \epsilon_{cs}) + \\ U(L_c - r_s - r_{cs}, \epsilon_{cs}) \Big],
\end{multline}
where $r_s$ represents the distance between the particles in the substrate and $r_{cs}$ the substrate-catalyst particle distance. The calculations for $\Delta U_{C{:}S \to C{:}P}$ and $\Delta U_{C:S \to C\cdot S}$ assume that $C{:}S$ is strictly one dimensional, which we have shown is true within the range of $L_c$ values necessary for catalysis (Fig.~S2B).

\textit{Diffusive processes}. Transitions $C+S \to C{\cdot}S$, $C{\cdot}S \to C{:}S$ and $C{:}P\to C{:}S$ are limited by diffusion. While the first transition depends on the volume of the system, the two latter are intrinsic to the catalyst design and only depend on $L_c$. We verify that in the narrow escape limit, i.e., $L_c/3r_{\min} > 1$, the first-passage time distributions for $C{\cdot}S \to C{:}S$ and $C{:}P \to C{:}S$ are described by a single timescale (Fig.~S4). To interpolate between the reversible and irreversible cases, we introduce a parameter $\gamma$, such that $ \tilde{k}_{C{:}P \to C{:}S} = \gamma k_{C{:}P \to C{:}S} $, where $\gamma = 1$ accounts for the reversible (diffusion-limited) case and $\gamma = 0$ accounts for the irreversible case. The first-passage time distribution associated with the $C+S\to C{\cdot}S$ transition is not exponential and depends on the initial distance of the substrate with respect to the catalyst~\cite{Grebenkov2019}. To estimate a rate, we map the transition to a search process in a disk of radius $R_{\text{Diff}}$ with reflecting boundary and an absorbing trap with radius $r = r_{\text{cutoff}}$ in the center. To compute the mean first-passage time to reach the absorbing trap starting from the reflecting boundary, we solve $D\Delta t(r) = -1$, where $D$ is the diffusion constant, with boundary conditions $\nabla t(r)|_{r= R_{\text{Diff}}} = 0 $ and $ t(r = r_{\text{cutoff}}) = 0$. We use the inverse of the mean first-passage time $k_{C+S \to C{\cdot}S} = 1/t(R_{\text{Diff}})$ as the rate. We note that the necessary and sufficient conditions for catalysis do not depend on this particular transition (SI), and hence, the value we set for this rate only impacts the catalyst's efficiency and not the regions where catalysis emerges. In particular, the maximal catalytic efficiency, depicted in Fig~3B in the main text, is obtained in the limit when $k_{C+S\to C{\cdot}S} \gg 1$. To account for the diffusion of the monomers towards and from the catalyst (Fig.~\ref{fig:efficiency}C and D), we extend the MSM in Fig.~\ref{fig:efficiency}A. See section 6 in SI for further details.

\bibliography{catbiblio}

\pagebreak
\onecolumngrid
\begin{center}
\textbf{\large Catalysis from the bottom-up: Supplementary Information} \\
Maitane Muñoz-Basagoiti, Olivier Rivoire, Zorana Zeravcic
\end{center}

\setcounter{equation}{0}
\setcounter{figure}{0}
\setcounter{table}{0}
\setcounter{page}{1}
\setcounter{section}{0}
\makeatletter
\renewcommand{\theequation}{S\arabic{equation}}
\renewcommand{\thefigure}{S\arabic{figure}}

\section{Mean reaction time for the spontaneous dissociation}\label{spontaneous}
\begin{figure}[b!]
    \centering
    \includegraphics[width =0.8\textwidth]{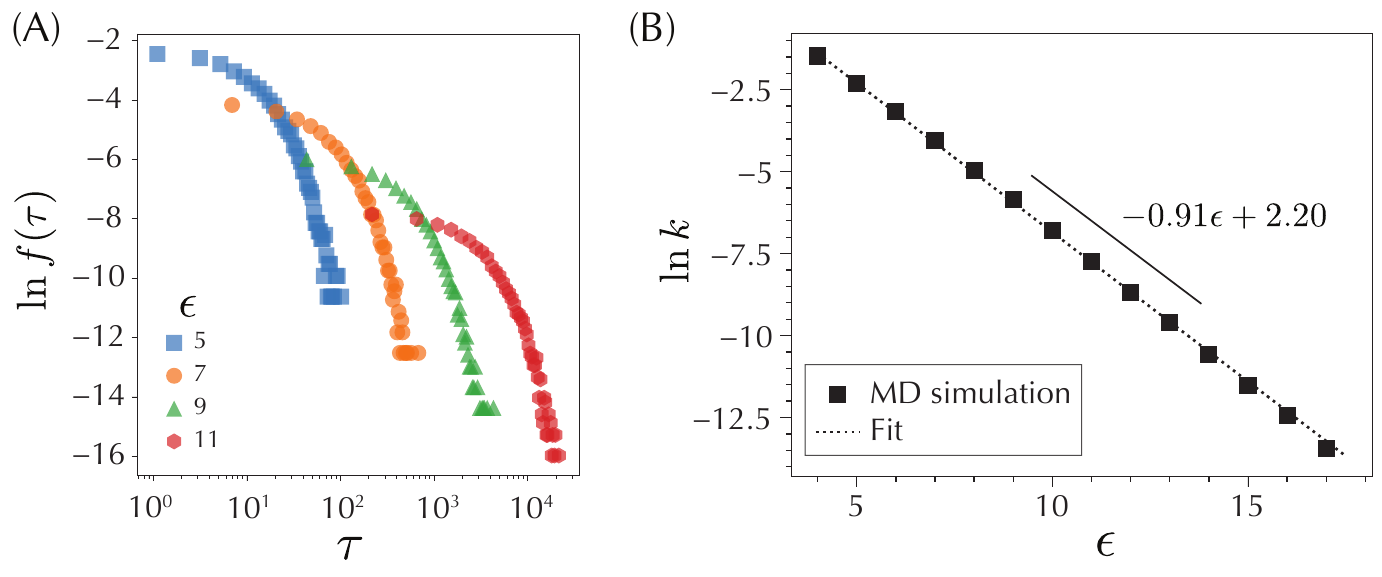}
    \caption{(A) Distribution of first-dissociation times $\tau$ obtained from MD simulations for different potential depths $\epsilon$, with fixed interaction range $r_{\text{rcutoff}}/\sigma = 1.1$. The dissociation occurs when the distance between particles in the substrate dimer, initially at equilibrium $r = r_{\min}$, exceeds $r>r_{\text{cutoff}}$ for the first time. (B) Logarithm of the dissociation rate $k$, defined as the inverse of the mean first-dissociation time, $k = 1/\left\langle \tau \right\rangle$, as a function of $\epsilon$. Simulation data (squares) can be fit with an Arrhenius-like expression $\ln k = -0.91\epsilon + 2.20$ (dashed line).}
    \label{fig:spontaneous}
\end{figure}

Using Molecular Dynamics (MD) simulations, we verify that the time it takes the substrate dimer to spontaneously dissociate into two free product monomers for the first time, that is, the first-passage time to overcome the interaction potential (Methods), is exponentially distributed (Fig.~\ref{fig:spontaneous}A). For fixed interaction range $r_{\text{cutoff}}/\sigma = 1.1$, the rate constant associated with the spontaneous reaction, defined as the inverse of the mean first-dissociation time~\cite{kramers1940brownian, Muller1997}, $k_{S\to P} = 1/T_{S\to P}$, decreases exponentially with the depth of the potential $\epsilon$ (Fig.~\ref{fig:spontaneous}B). We fit the simulation data with an Arrhenius-like expression $k = e^{-A\epsilon +B}$, with $A = 0.91$ and $B = 2.20 \pm 0.07$ within the range $\epsilon \in [3, 17]$. Since $A < 1$, the $\epsilon_{cs} < A\epsilon_s < \epsilon_s$ necessary condition for catalysis holds (Methods).

\section{A spherical particle cannot catalyze the dissociation reaction}\label{sphericalparticle}
A single spherical particle, i.e. $L_c =0$, cannot catalyze a dissociation reaction within our model. This is shown in Fig.~\ref{fig:geometry}A, where we plot the free energy of the system along the substrate bond $r_s$ when the substrate is completely attached to the catalyst, i.e., $r_{cs}<r_{\text{cutoff}}$ for the two substrate-catalyst bonds. We compare two different catalyst designs: a single catalyst particle and a rigid dimer with $L_c > 3r_{\min}$. The activation barrier to break the substrate bond in the presence of a single catalyst particle with $\epsilon_{cs}/\epsilon_s = 1$ equals the barrier in the absence of the catalyst. In other words, the interaction between the substrate dimer and the catalyst particle does not modify the average time to cleave the substrate bond. For the same binding energy, a rigid dimer satisfying $L_c>3r_{\min}$ reduces the activation barrier by straining the substrate bond, consequently accelerating the cleaving of the bond.
 
\section{The \texorpdfstring{$L_c>3r_{\text{min}}$}{Lc>3rmin} necessary condition for the catalyst geometry}
\begin{figure}[t!]
    \centering
    \includegraphics[width = \textwidth]{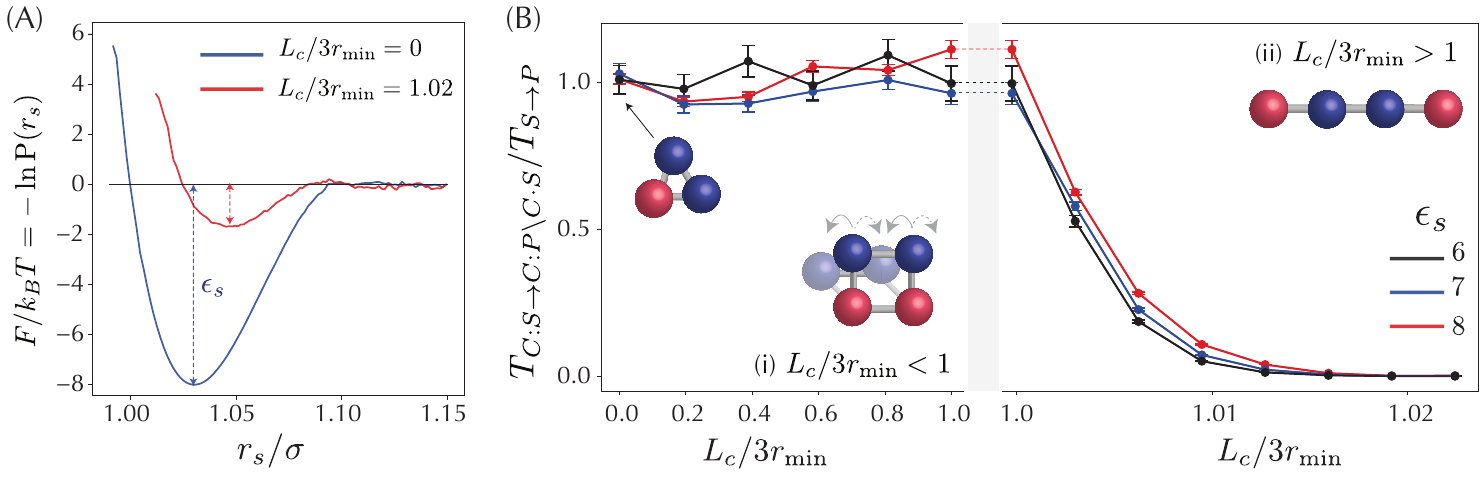}
    \caption{(A) Free energy of the system along the substrate bond $r_s$ when the substrate is fully attached to the catalyst ($r_{cs}<r_{\text{cutoff}}$ for both substrate-catalyst bonds) for two catalyst designs: a single spherical particle (blue line) and a rigid dimer with $L_c/3r_{\min} = 1.02$ (red line). Simulation parameters correspond to $\epsilon_s = \epsilon_{cs}= 8$. (B) Comparison of the mean first-passage times to break the substrate bond with and without the catalyst, $T_{C{:}S \to C{:}P \backslash C{\cdot}S}/T_{S \to P}$, as a function of $L_c$, where we set $\epsilon_{cs} = 30 \gg \epsilon_s$ to avoid unbinding events. Catalysis requires $T_{C{:}S \to C{:}P \backslash C{\cdot}S }/T_{S \to P}<1$. When $L_c\leq 3r_{\min}$, all three bonds in the $C{:}S$ configuration can exist at equilibrium. This is no longer true when $L_c>3r_{\min}$, and as a result, the substrate bond is strained by the catalyst.}
    \label{fig:geometry}
\end{figure}

Catalysis requires $T_{C{:}S \to C{:}P \backslash C{\cdot}S} < T_{S\to P}$, where $T_{C{:}S \to C{:}P \backslash C{\cdot}S}$ is a function of the catalyst geometry $L_c$, the $\epsilon_{cs}$ binding energy and the substrate bond $\epsilon_s$. To investigate the effect of $L_c$ on $T_{C{:}S \to C{:}P \backslash C{\cdot}S}$, we initiate the system in $C{:}S$ and record the time it takes to reach $C{:}P$ using MD simulations (Methods). To prevent unbinding events, we set $\epsilon_{cs}\gg \epsilon_{s}$. Simulation results in Fig.~\ref{fig:geometry}B show that catalysis requires $L_c/3r_{\min}>1$. In this limit, the catalyst geometry does not allow the three bonds in $C{:}S$, namely, the substrate bond and the two catalyst-substrate bonds, to simultaneously exist at equilibrium. As a consequence, the substrate bond can be strained and the cleaving of the bond accelerated. Note that despite the large interaction energy $\epsilon_{cs}$ available, the catalyst only accelerates the reaction for a small subset of geometries, in agreement with Pauling's principle of transition state stabilization~\cite{Pauling1946}. For this subset of geometries, the isotropic interaction with the catalyst becomes directional.
 
\begin{figure}[t]
    \centering
    \includegraphics[width=0.8\textwidth]{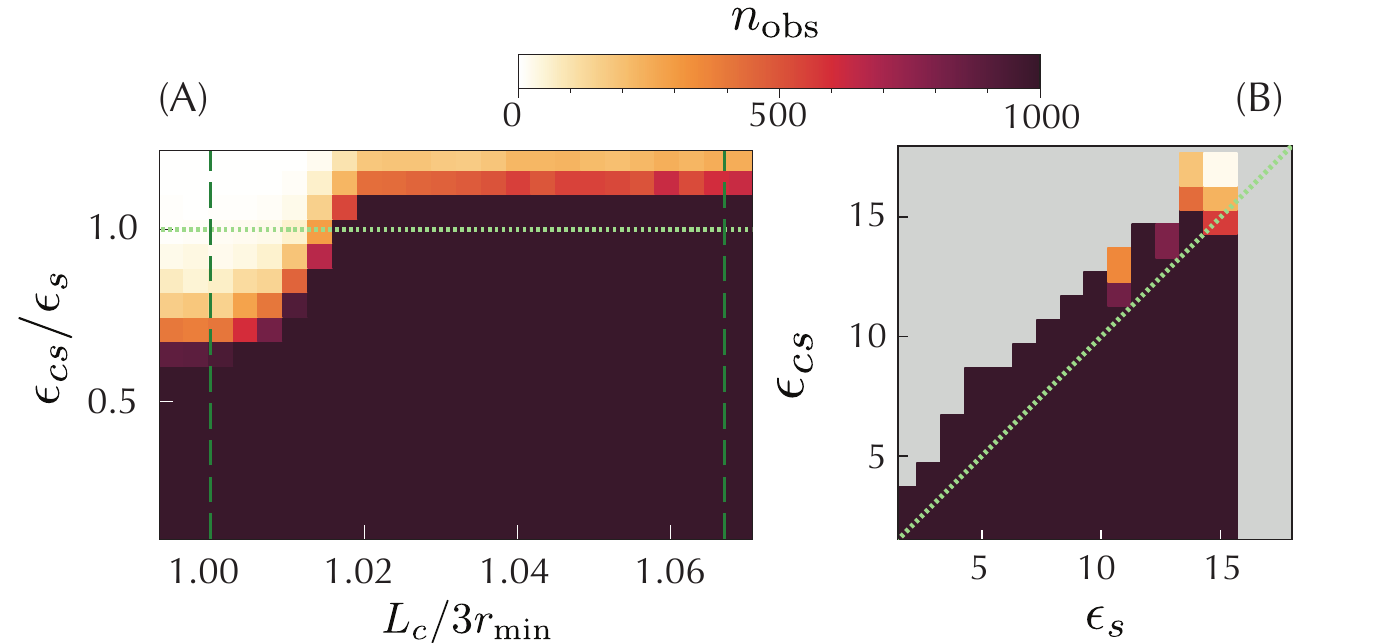}
    \caption{Number of observations $n_{\text{obs}}$ for each data point reported in Fig~2 in the main text. The color bar is capped at $n_{\text{obs}} =1000$ observations. The gray area in panel (B) indicates that there is no data in this region. The $\epsilon_{cs} < \epsilon_s$ (green dotted lines) and $3 r_{\min} < L_c < 3r_{\text{cutoff}}$ (gree dashed lines) conditions for catalysis are indicated.}
    \label{fig:stats}
\end{figure}
\section{Simulation data and statistics for Fig.~2}\label{section:fig3}

To produce Fig.~2 in the main text, we initiate an ensemble of simulations in $C+S$ by placing the substrate at a random $r>r_{\text{cutoff}}$ distance from both particles in the catalyst in a simulation box with side $L/\sigma = 7.5$, and record the time it takes two monomers to be released in solution in the presence of the catalyst. This time is compared to the timescale of the spontaneous reaction (see SI section~\ref{spontaneous}). Each time a monomer is released in solution, it is systematically blocked, that is, it is not allowed to interact with any other particle. We only perform simulations for the reversible case, where the substrate bond can reform in the presence of the catalyst through the $C{:}P \to C{:}S$ transition. To account for the irreversible case, which concerns Fig.~2 in the main text, we use the same simulation trajectories, but only keep the trajectory up to the first time the substrate bond breaks. If the bond breaks after a $C{:}S \to C{:}P$ or a $C{\cdot}S \to C{\cdot}P$ transition, we then simulate the release of the monomer(s) bound to the catalyst by drawing from an exponential distribution with the rate $k(\epsilon_{cs})$ ( Fig.~\ref{fig:spontaneous}B). The number of simulations run for each set of parameters is shown in  Fig.~\ref{fig:stats}.

\begin{figure}[t]
    \centering
    \includegraphics[width=0.7\textwidth]{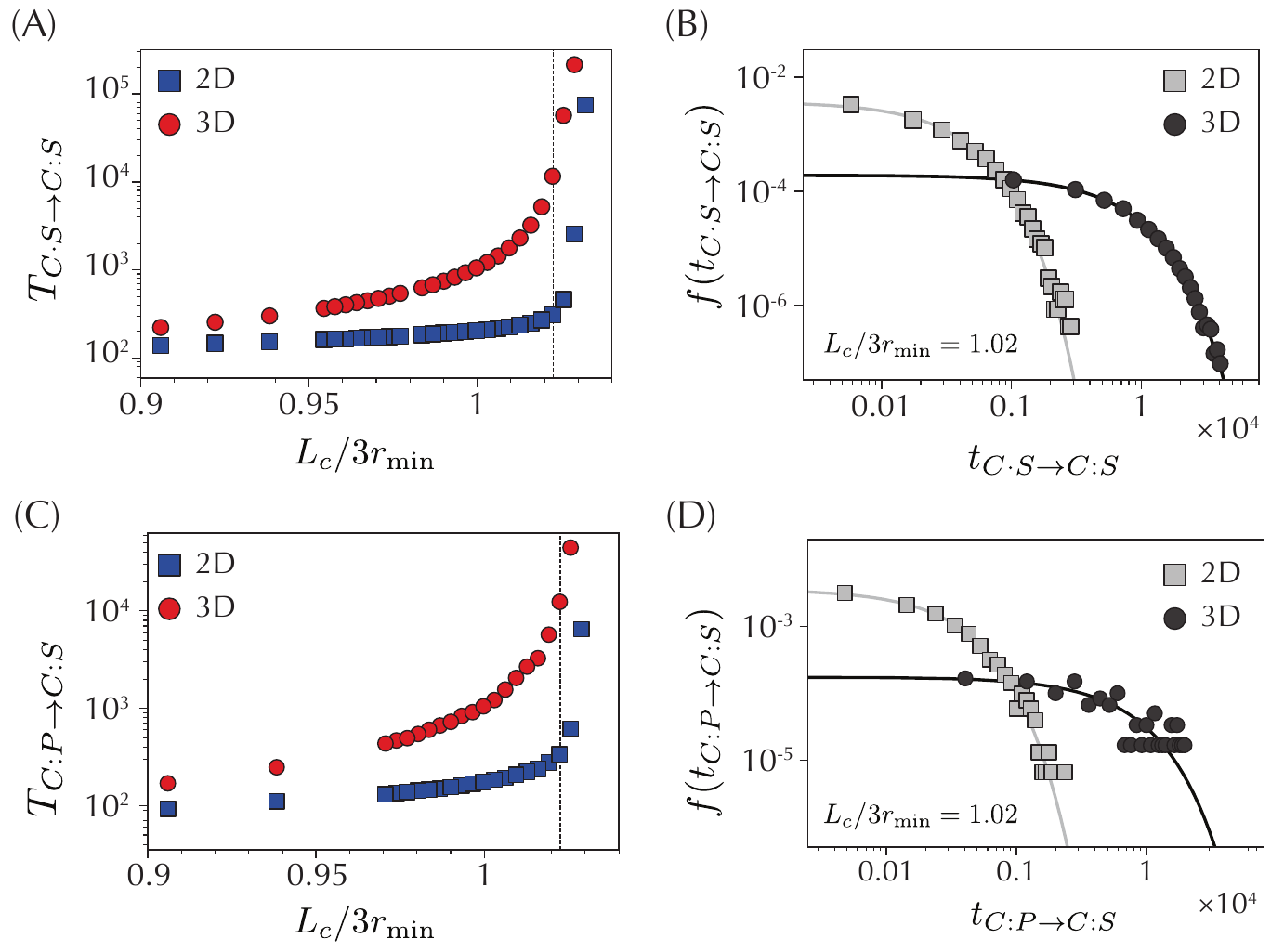}
    \caption{(A, C) Simulated mean first-passage time for the $C{\cdot}S\to C{:}S$ (A) and $C{:}P \to C{:}S$ (C) transitions as a function of $L_c$ for a system in 2D (blue squares) and 3D (red circles). The vertical dashed line indicates the $L_c = 2r_{\min}+r_c$ threshold. (B, D) First-passage time distributions for $C{\cdot}S\to C{:}S$ (B) and $C{:}P \to C{:}S$ (D) when $L_c/3r_{\min}=1.02$, for simulations in two (gray squares) and three (black circles) dimensions. The solid lines are exponential distributions with rate $k = 1/T_{C{\cdot}S \to C{:}S}$ and $k = 1/T_{C{:}P \to C{:}S}$ taken from (A) and (C), respectively, for the corresponding geometry.}
    \label{fig:binding}
\end{figure}

\section{First-passage times to bind and reform the substrate bond}
The $C{\cdot}S \to C{:}S$ transition is diffusion-limited and only depends on the catalyst geometry. To measure the mean first-binding time associated to it, we initiate the system in a random configuration corresponding to state $C{\cdot}S$ and record the time its takes to reach $C{:}S$ for the first time, i.e., the time for the free particle in the substrate to attach to the catalyst. The average first-binding time diverges as $L_c \to 3r_{\text{cutoff}}$, as shown for a system in 2D and 3D in  Fig.~\ref{fig:binding}A. This gives rise to the $L_c/3r_{\min} < r_{\text{cutoff}}/r_{\min}$ necessary condition for catalysis and impacts the trade-off for optimal catalyst geometry. Indeed, the further apart the two particles in the catalyst, the longer it will take the substrate to fully bind. We note that when $L_c/ 3r_{\min} \gg 1$ (narrow escape limit), the first-binding time distribution is dominated by an exponential ( Fig.~\ref{fig:binding}B). 

The $C{:}P \to C{:}S$ transition is similar to the $C{\cdot} S \to C{:}S$ process: it is also limited by diffusion and only depends on the catalyst geometry $L_c$. We provide the average first-time to reform the substrate bond, and first-passage time distribution in Fig.~\ref{fig:binding}C and D. These are obtained by initiating the simulations in $C{:}P$ and recording the time it takes to reform the substrate bond, i.e., reach $C{:}S$ for the first time.

\section{Markov State Models}\label{sectionMSM}
We use Markov State Models (MSMs) to extend our results beyond the parameter range that we can explore with MD simulations. We first construct a minimal model to account for the system when free monomers are systematically removed once they are released in solution. The structure of this model is shown in Fig.~3A in the main text, and it is used to produce Fig~3B (black lines). We then extend the model to account for the system when free monomers are removed only if they diffuse sufficiently far from the catalyst, i.e., $r>R_{\text{Diff}}$. This extended model is used to produce the results in Fig~3C and Fig.~3D in the main text. 

\begin{figure}[t!]
    \centering
    \includegraphics[width =\textwidth]{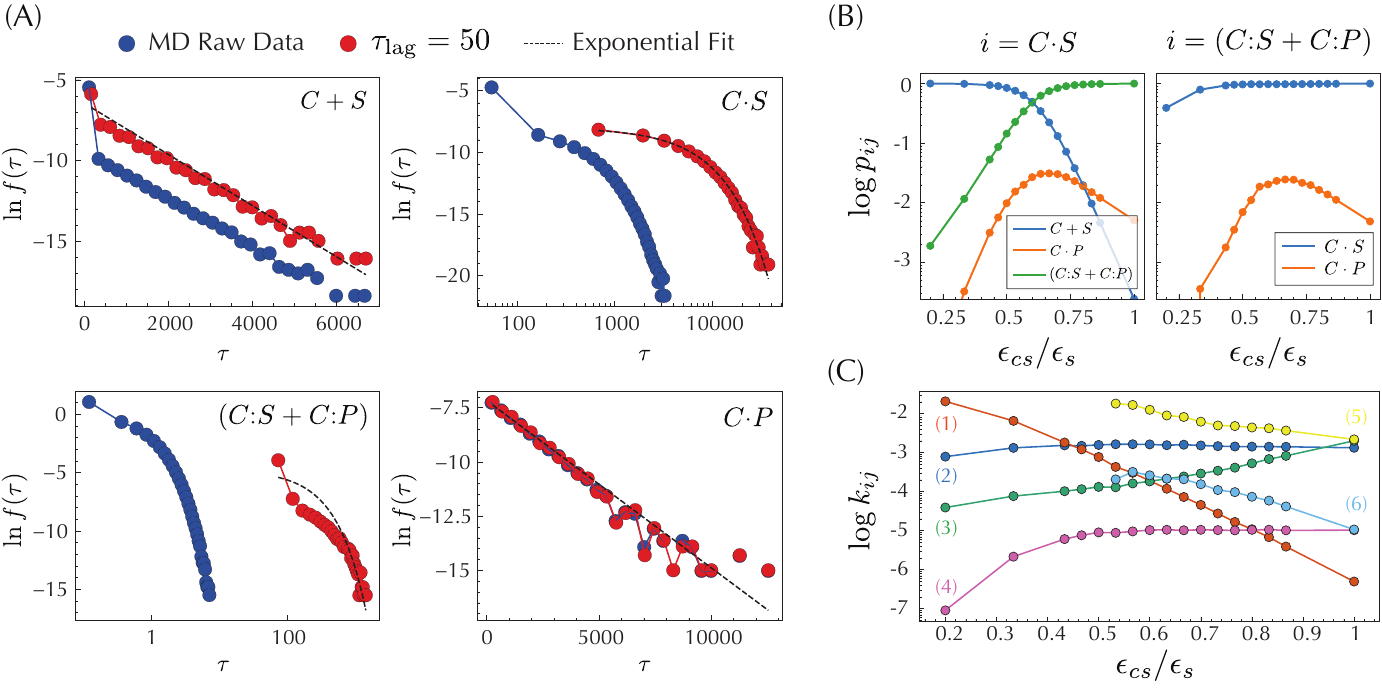}
    \caption{(A) Waiting time distributions for states $C+S$, $C{\cdot}S$, $C{\cdot}P$ and the merged state $(C{:}S+C{:}P)$ (Methods), obtained by recording the time it takes the system to form and break bonds (blue) or by coarse-graining the MD trajectories with a lag time $\tau_{\text{Lag}} = 50 $ time units, which allows us to fit an exponential distribution to the data (dashed line) and to extract the average waiting time $\left\langle \tau_i \right\rangle$. Simulation parameters are $\epsilon_s = 15$ and $\epsilon_{cs} = 10.5$. Results correspond to the reversible limit. (B) Transition probabilities $p_{ij}$ from states $i = C{\cdot}S$ and $i = (C{:}S+C{:}P)$ as a function of the catalyst binding strength $\epsilon_{cs}$. Legend indicates states $j$ where the system transitions next. (C) Transition rates $k_{ij}=p_{ij}/\left\langle \tau_i \right\rangle$ as a function of the catalyst binding energy $\epsilon_{cs}$: (1) $k_{C{\cdot}S \to C+S}$, (2) $k_{C+S \to C{\cdot}S}$, (3) $k_{C{\cdot}S \to (C{:}S + C{:}P)}$, (4) $k_{C{\cdot}S \to C{\cdot} P}$, (5)  $k_{(C{:}S + C{:}P) \to C{\cdot}S}$ and (6) $k_{(C{:}S + C{:}P) \to C{\cdot}P}$.}
    \label{fig:MSM}
\end{figure}

\subsection{Minimal MSM}
The minimal MSM, depicted in Fig~3A in the main text, consists of six states, $C+S$, $C{\cdot}S$, $C{:}S$, $C{:}P$, $C{\cdot}P$ and $C+P$, where each state is characterized by the number and the types of bonds (see configurations in Fig~1C in the main text). We first extract rates from numerical simulations (Methods). Examples of waiting time distributions, transition probabilities and simulation-inferred rates are shown in Fig.~\ref{fig:MSM}. Next we derive analytical expressions for the dependence of the rates on the parameters of the system. This allows us to explore the parameter space beyond simulations (Methods).

\begin{figure}[t]
    \centering
    \includegraphics[width=\textwidth]{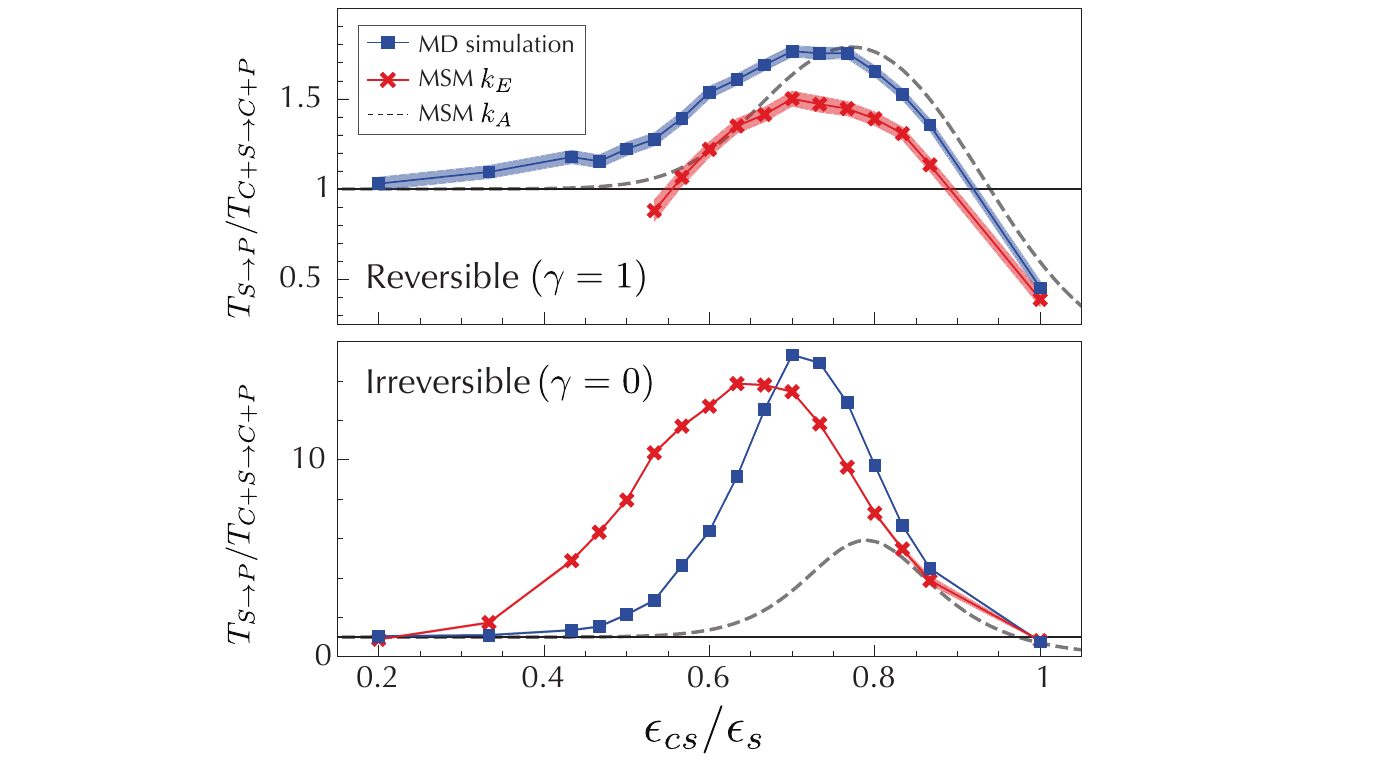}
    \caption{Comparison of the catalytic efficiency, i.e., $T_{S\to P}/T_{C+S \to C+P}$, for a fixed substrate bond ($\epsilon_s = 15$) and catalyst geometry ($L_c /3r_{\min} = 1.02$) and different $\epsilon_{cs}$ for numerical simulations (blue points), the MSM with the rates inferred from simulation (red crosses, $k_E$) and the MSM using the analytical rates (dashed black line, $k_A$) (Methods). Top panel corresponds to the reversible case ($\gamma = 1$), while the bottom panel corresponds to the irreversible case ($\gamma = 0)$. In all cases, monomers are removed as soon as they are released in the solution. Simulation results correspond to a box with side $L/\sigma = 15$. }
    \label{fig:comparisonvalidation}
\end{figure}

To validate the coarse-graining and discretization of the catalytic path into states, we compare the efficiency of the catalyst design, i.e., $T_{S\to P}/T_{C+S \to C+P}$, in simulations, where the catalytic pathway is \textit{not} divided into elementary transitions, to the efficiency using the minimal MSM with rates inferred from simulations and rates from the analytical model. The comparison is shown in  Fig.~\ref{fig:comparisonvalidation} for the reversible ($\gamma = 1$) and irreversible ($\gamma = 0$) cases. The agreement in the reversible case supports the coarse-graining of the system into states.  The way in which data is generated for the irreversible case (see SI section~\ref{section:fig3}) might be one of the reasons why the results do not quantitatively agree for small $\epsilon_{cs}$ for the MSM with the simulation inferred rates. Although the analytical model for the rates underestimates the efficiency of the catalyst in the irreversible case, the overall scaling agreement between the MSM and simulation data in Fig.~3B in the main text for both reversible and irreversible cases further support the coarse-graining.

\subsection{Extended MSM to account for volume}
\begin{figure}[t!]
    \centering
    \includegraphics[width=\textwidth]{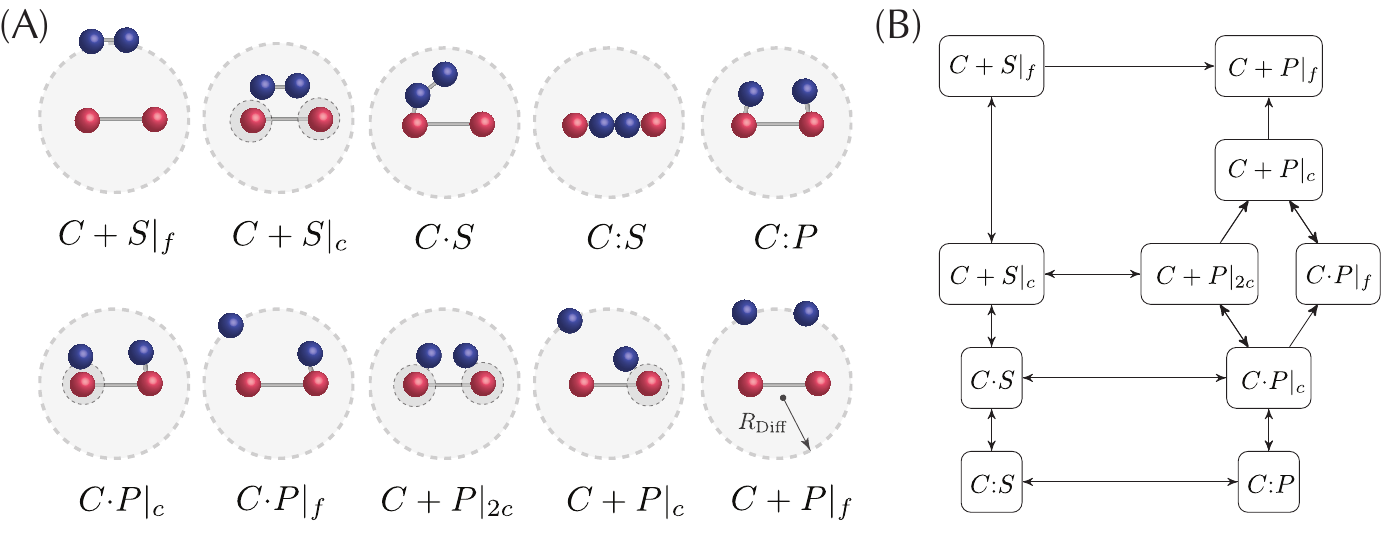}
    \caption{(A) States considered in the extended MSM, when product monomers (blue) are not immediately removed after they are released in solution and have to diffuse a distance $r>R_{\text{Diff}}$ from the center of the system, where the catalyst is placed (red), to leave it. We now consider that the substrate and product particles particles can be far ($f$) or close ($c$) to the catalyst. (B) Extension of the MSM in Fig.~3A in the main text to account for the diffusion of the monomers away from the catalyst.}
    \label{fig:configurationsvolume}
\end{figure}

To explore the impact of product inhibition and other volume effects on catalysis, we require monomers to diffuse sufficiently far away from the catalyst, i.e., $r>R_{\text{Diff}}$, in order to leave the system, where $R_{\text{Diff}}$ is the radius of a disk (sphere) centered around the catalyst (see  Fig.~\ref{fig:configurationsvolume}A). We extend the MSM in Fig~3A in the main text to 10 states and introduce transitions that describe the substrate and product particles diffusing towards or away from the catalyst. To estimate a rate for these transitions, we take the inverse of the mean first-passage time of a search process in a bounded domain. We compute the mean first-passage time by solving $D\Delta t(r) = -1 $ with the appropriate boundary conditions, where $D$ is the diffusion constant~\cite{Redner.2001}. For example, we estimate the mean first-passage time for the $C+S|_f \to C+S|_c$ transition, where the subscripts $f$ and $c$ indicate that the substrate is 'far' ($r =R_{\text{Diff}}$) and 'close' ($r = r_{\text{cutoff}} + \delta $, with $\delta \ll 1$) from the catalyst, by mapping the transition to a diffusion process in a disk (sphere) with boundary conditions $
t(r= r_{\text{cutoff}} +\delta) = 0 $ and $\nabla t(r)|_{ r = R_{\text{Diff}}} = 0 $. The first condition represents an absorbing target of radius $r = r_{\text{cutoff}} + \delta$ in the center of the domain and the second one accounts for the reflecting boundary at $r = R_{\text{Diff}}$. The mean first-passage time will depend on the starting point of the process $t = t(r_0)$, which we set at $r_0 = R_{\text{Diff}}$, corresponding to state $C+S|_f$. We estimate the first return time for $C+S|_c \to C{\cdot}S$ by assuming a similar setup, but setting the starting point of the process at $r_0 = r_{\text{cutoff}} + \delta$ and absorbing boundary at $r = r_{\text{cutoff}}$. Transitions $C{\cdot}P|_c \to C{:}P$, $C+P|_{2c}\to C{\cdot}P|_c$ and $C+P|_c \to C{\cdot}P|_f$ also fall within this category. We note that results depend on the choice of $\delta$, and we use $\delta = 0.01$ to produce Fig.~3C and D in the main text. 

For transitions requiring the monomers to diffuse away from the catalyst, such as $C{\cdot}P|_c \to C{\cdot}P|_f$ and $C+P|_{2c}\to C+P|_c$, we consider $t(r = R_{\text{Diff}}) = 0 $ and $\nabla t(r)|_{ r = r_{\text{cutoff}}} = 0 $ as boundary conditions,
and set the starting point of the process at $r_0 = r_{\text{cutoff}}$. We note that despite our procedure to estimate the rates, the first-passage time distributions for these processes are not exponential~\cite{MejiaMonasterio2011, Grebenkov2019}, and therefore, a rate cannot be properly defined. The comparison between MD simulation data in Fig.~3C in the main text, and  Fig.~\ref{fig:my_label} below show qualitative agreement. Further improvements of the model regarding diffusive processes are left as future work.

\section{Necessary conditions for catalysis.}\label{conditions}
We derive the necessary conditions for catalysis when monomers are systematically removed from the system (MSM in Fig.~3A in the main text) by comparing the mean first-passage time from $C+S$ to $C+P$, i.e., $T_{C+S \to C+P}$, to the mean reaction time in the absence of catalyst, $T_{S\to P} = 1/k_{S\to P}$. $T_{C+S \to C+P}$ can be analytically computed~\cite{IyerBiswas} and the criterion for catalysis, $T_{S\to P}/T_{C+S \to C+P}> 1$, takes the form
\begin{multline}\label{eq:schememodel}
      \frac{1}{k_{S\to P}} > \frac{1}{k_{C{:}S\to C{:}P}} +  \frac{1}{k_{C{:}P \to C{\cdot}P}} +  \frac{1}{k_{C{\cdot}P \to C+P}} \\  +  \frac{k_{C{:}P\to C {:}S}}{k_{C{:}S\to C{:}P} k_{C{:}P \to C{\cdot}P}} +   \frac{k_{S\to P}}{k_{C{\cdot}S \to C{:}S} k_{C{\cdot}P \to C+P}}  \times \left(1 + \frac{k_{C{:}S \to C{\cdot}S} ~k_{C{:}P\to C {:}S}}{ k_{C{:}S\to C {:}P} ~k_{C{:}P \to C{\cdot}P}} +  \frac{k_{C{:}S\to C{\cdot}S}}{k_{C{:}S \to C{:}P} } \right),
\end{multline}
where $k_{i\to j}$ is the rate to transition from state $i$ to state $j$. The above equation is a sufficient condition for catalysis of the form $T_{S\to P} > \sum_n T_n$.  Each individual term on the right hand side of the equation leads to a necessary condition for catalysis that can be subsequently translated into physical and geometrical constraints in our model. 

The first necessary condition in eq.~\eqref{eq:schememodel}, $k_{S \to P} < k_{C{:}S\to C{:}P}$, pertains to the catalytic mechanism and leads to $L_c > 3r_{\min}$ and $\epsilon_{cs} > 0$. The second and third necessary conditions in eq.~\eqref{eq:schememodel}, $k_{S\to P} < k_{C{:}P \to C{\cdot}P}$ and $k_{S \to P} < k_{C{\cdot}P \to C+P}$, are associated to product release and lead to $\epsilon_{cs} < \epsilon_s$. 
The fourth necessary condition in eq.~\eqref{eq:schememodel} is non trivial only in the reversible case, when $C{:}P \to C{:}S$ is possible. In the irreversible case, this transition is not allowed and hence, $k_{C{:}P\to C {:}S} = 0$. The last necessary condition on the right hand side in equation~\eqref{eq:schememodel} stems from the alternative pathway that the substrate may take in the presence of the catalyst to produce products, i.e. $C{\cdot} S \to C+P$, without visiting $C{:}S$. This shortcut to the final state of the reaction relaxes the constraint on $k_{C{\cdot}S \to C{:}S}$, which does not have to be as fast as the spontaneous reaction for catalysis to emerge.

Note that eq.~\eqref{eq:schememodel} is independent of $T_{C+S \to C{\cdot}S} $, the average time it takes the substrate to diffuse to the catalyst. This is because the substrate can always dissociate spontaneously into two monomers in a single step in the presence of the catalyst. As here we consider that the reaction ends when two monomers have been released into solution (systematic product removal), any catalyst design that successfully accelerates the production of monomers will only contribute to reducing the mean reaction time. In other words, as long as the product is systematically removed, catalysis is an intrinsic property of our catalyst design, and therefore independent of the volume of the system. The efficiency of the catalyst, however, depends on volume.

\section{Scaling of the maximal efficiency}
\begin{figure}
    \centering
    \includegraphics[width=0.8\textwidth]{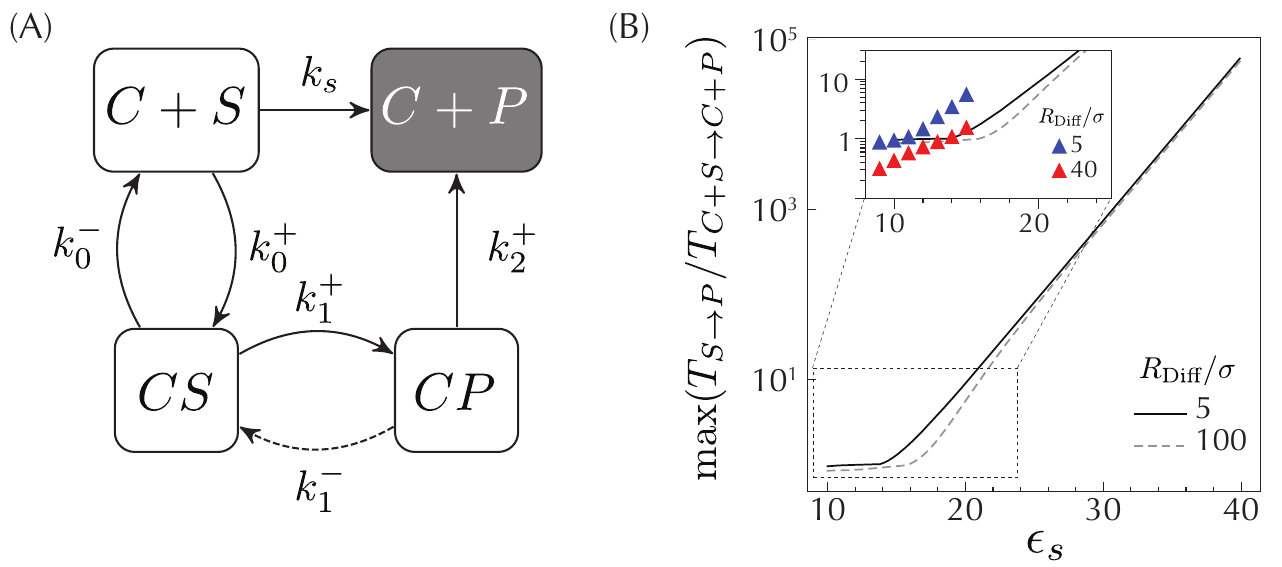}
    \caption{(A) Minimal reaction scheme that we use to explain the scaling of the maximal catalytic efficiency in the reversible and irreversible limits. (B) MSM model results for the maximal catalytic efficiency when $\gamma = 0$ and monomers are removed from the system if $r>R_{\text{Diff}}$, for $R_{\text{Diff}}/\sigma = 5$ (black solid line) and $R_{\text{Diff}}/\sigma = 100$. (Inset) Zoom in the small $\epsilon_s$ region showing MD simulation data for $R_{\text{Diff}} /\sigma= 5$ (blue triangles) and $R_{\text{Diff}} /\sigma= 40$ (red triangles).}
    \label{fig:minimalscheme}
\end{figure}

To understand why the maximal catalytic efficiency scales exponentially with $\epsilon_s$ in the irreversible case and why it saturates in the reversible case (Fig~3B in the main text), we propose a minimal reaction scheme that captures the essential features of our model. This scheme is shown in  Fig.~\ref{fig:minimalscheme} and it contains four states, where the $CS \to CP$ transition accounts for the chemical transformation step and $CP\to C+P$ represents product release. 
We model the rates as
\begin{equation}
    k_s = e^{-\epsilon_s} \qquad k_1^+ = k_s e^{\alpha \epsilon_{cs}} \qquad k_2^+ = e^{-\epsilon_{cs}} \qquad k_1^- = \gamma k_{CP \to CS},
\end{equation}
and we leave $k_0^+$ and $k_0^-$ unspecified as they do not take part in the sufficient condition for catalysis (see SI section~\ref{conditions}). Here $\alpha$ represents the catalyst's ability to reduce the rate of the spontaneous reaction, $k_{CP\to CS}$ is the diffusion-limited rate of the reverse reaction in the catalyst and $\gamma$ is the parameter that regulates such rate. Note that for fixed $\epsilon_s$, the larger $\epsilon_{cs}$, the larger $k_1^+$ will be, but the smaller $k_2^+$, recovering the Sabatier principle of optimal intermediate binding strength.  To make the notation compact, we use $\gamma k_{CP \to CS} \equiv \tilde{\gamma} $. Catalysis requires $\alpha>0$ and $\epsilon_{cs} < \epsilon_s$. When $\alpha = 1$, the catalyst cancels the barrier of the spontaneous reaction.

We determine the efficiency of the catalyst by computing the mean first-passage time to reach the product state in the presence and absence of the catalyst, which yields
\begin{equation}
    \frac{T_{S\to P}}{T_{C+S \to C+P}} = \frac{k_0^+ k_1^+ k_2^+ + k_1^+ k_2^+ k_s + k_0^- (k_1^- + k_2^+)k_s}{[k_1^+ k_2^+ + k_0^- (k_1^- + k_2^+) + k_0^+ (k_1^- +k_1^+ + k_2^+)] k_s}.
\end{equation}

\subsection{Irreversible case: exponential scaling}
In the limit when $k_0^+ \to \infty$ and $\tilde{\gamma} = 0$, the mean-first passage time to produce two product monomers in the presence of the catalyst is
\begin{equation}
    T_{C+S \to C+P} = \frac{1}{k_1^+} + \frac{1}{k_2^+} = e^{\epsilon_{cs}} + e^{\epsilon_{s}-\alpha\epsilon_{cs}},
\end{equation}
which is minimized by
\begin{equation}
    \epsilon_{cs}^* = \frac{\epsilon_s + \log \alpha}{1+\alpha}.
\end{equation}
As a result, the maximal catalytic efficiency for $\epsilon_s \to \infty$ scales as
\begin{equation}
  \lim_{\epsilon_s \to \infty} \left(\frac{T_{S\to P}}{T_{C+S \to C+P}} \right) \Big|_{\epsilon_{cs} = \epsilon_{cs}^*} = \frac{1}{1+\alpha} e^{\frac{1}{1+\alpha} (\epsilon_s + \log \alpha)} \sim e^{\tilde{\alpha} \epsilon_s},
\end{equation}
recovering the scaling in Fig. 3B in the main text for the irreversible case, where $\tilde{\alpha} = (1+\alpha)^{-1}$. Note that when $\alpha = 1$, the optimal binding strength $\epsilon_{cs}^* = \epsilon_s/2$ and $\tilde{\alpha} = 0.5$. For our MD results in Fig~3B, $\alpha_{MD} = 0.50\pm 0.04$.

\subsection{Reversible case: Saturation}\label{saturation}
\noindent When $\tilde{\gamma} \neq 0$, $T_{C+S \to C+P}$ in the limit when $k_0^+ \to \infty$ is given by

\begin{equation}
    T_{C+S \to C+P } = \frac{1}{k_1^+} + \frac{1}{k_2^+} + \frac{k_1^-}{k_1^+ k_2^+} = e^{\epsilon_{cs}} + e^{\epsilon_{s}-\alpha\epsilon_{cs}} + \tilde{\gamma}e^{(1-\alpha)\epsilon_{cs} +\epsilon_s},
\end{equation}
and the catalytic efficiency is
\begin{equation}
    \frac{T_{S\to P}}{T_{C+S \to C+P}} = \frac{e^{\epsilon_s}} {e^{\epsilon_{cs}} + e^{\epsilon_{s}-\alpha\epsilon_{cs}} + \tilde{\gamma}e^{(1-\alpha)\epsilon_{cs} +\epsilon_s}} =  \frac{1}{ e^{\epsilon_{cs} - \epsilon_s} + e^{-\alpha \epsilon_{cs}} + \tilde{\gamma} e^{(1-\alpha) \epsilon_{cs} }}
\end{equation}
Since catalysis requires $\epsilon_s > \epsilon_{cs}$, in the limit when $\epsilon_s \to \infty$,
\begin{equation}
\frac{T_{S\to P}}{T_{C+S \to C+P}} = \frac{1}{e^{-\alpha \epsilon_{cs}} + \tilde{\gamma} e^{(1-\alpha) \epsilon_{cs} }}.
\end{equation}
The catalytic efficiency is maximal when the denominator is the smallest, which is minimized when $\epsilon_{cs} = 0$. As a result, the maximal efficiency saturates,
\begin{equation}
\frac{T_{S\to P}}{T_{C+S \to C+P}} = \frac{1}{e^{-\alpha \epsilon_{cs}} + \tilde{\gamma} e^{(1-\alpha) \epsilon_{cs} }}\Big|_{\epsilon_{cs} = 0} = \frac{1}{1+\tilde{\gamma}},
\end{equation}
in agreement with the cases for $\gamma \neq 0$ in Fig.~3B in the main text.

\section{Emergence of catalysis in 3D}
\begin{figure}[t!]
    \centering
    \includegraphics[width = 0.8\textwidth]{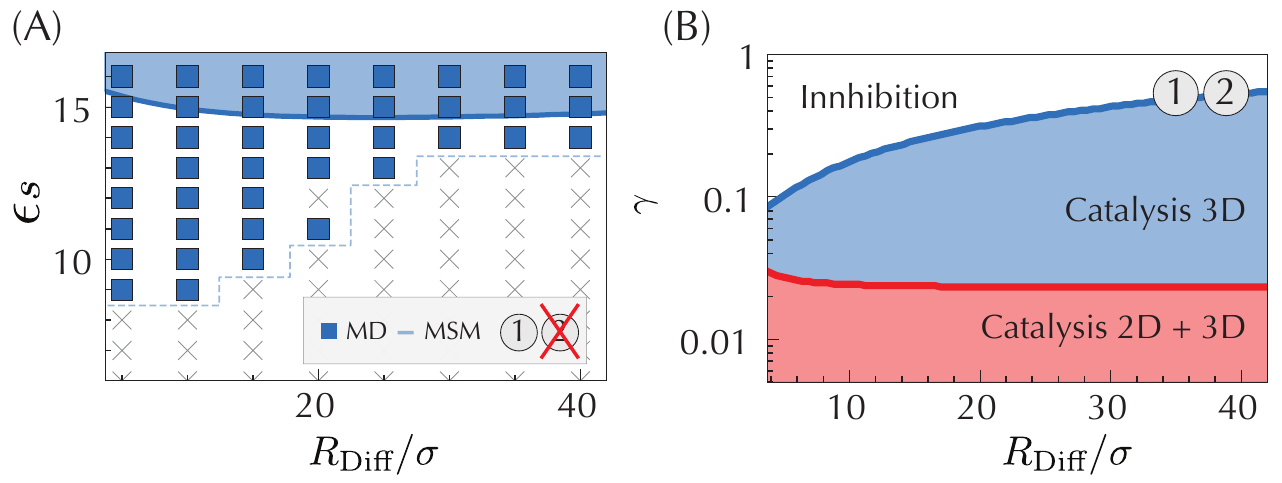}
    \caption{Simulation and MSM data for the system in 3D. (A) Substrate bonds $\epsilon_s$ for which there is catalysis when monomers are removed from the system if they diffuse sufficiently far from the catalyst, i.e., $r>R_{\text{Diff}}$. Simulation data is shown as squares and crosses and the MSM results are shown as a shaded blue region. (B) MSM results showing values of $\gamma$ for which catalysis can be observed in 2D (red region) and 3D (red and blue regions) when monomers are removed from the system if they have diffused a distance $R_{\text{Diff}}/\sigma$ from the catalyst.}
    \label{fig:my_label}
\end{figure}

In Fig.~\ref{fig:my_label}, we produce Figs.~3C and D in the main text for a system in 3D. To construct Fig.~\ref{fig:my_label}A, we simulate the system at varying $\epsilon_s$ and $R_{\text{Diff}}$, with fixed catalyst geometry $L_c/3r_{\min} = 1.02$ and we explore a range of $\epsilon_{cs}$ that satisfy $\epsilon_{cs} < \epsilon_s$. We consider that catalysis emerges for a given $\epsilon_s$ and $R_{\text{Diff}}$ if there is at least one $\epsilon_{cs}$ for which $T_{S\to P}/T_{C+S \to C+P} > 1$. We note that since our resolution in binding energy is $\Delta \epsilon_{cs}/k_B T = 1 $, we may miss designs for which the criterion for catalysis is satisfied. The results qualitatively agree with the MSM (blue region). MD results in 3D indicate that catalysis emerges at smaller values of $\epsilon_s$ than in the 2D case, i.e., $\epsilon_{s, \min}^{(3D)}<\epsilon_{s, \min}^{(2D)}$. In Fig.~\ref{fig:my_label}B, we show MSM results where catalysis was observed for a range of $\epsilon_s \in [2, 30]$.

\end{document}